\documentclass{article}
\usepackage{amsmath}
\usepackage{graphicx}

\def\be{\begin{equation}}
\def\ee{\end{equation}}

\def\bea{\begin{eqnarray}}
\def\eea{\end{eqnarray}}

\def\la{\lambda}
\def\GeV{{\rm\ GeV}}
\def\meV{{\rm\ meV}}

\begin{document}

\title{Meson exchange in lepton-nucleon scattering and proton radius puzzle}
\author{Dmitry~Borisyuk\\[2mm]
\it Bogolyubov Institute for Theoretical Physics,\\
\it 14-B Metrologicheskaya street, Kiev 03680, Ukraine}
\maketitle

\begin{abstract}
We study two-photon contribution to the elastic lepton-proton scattering, associated with $\sigma$ meson exchange, with special attention paid to the low-$Q^2$ region.
%The effect is proportional to the lepton mass, and is therefore much more pronounced for the muon than for the electron.
We show that the corresponding amplitude grows sharply but remains finite at $Q^2\to 0$.
The analytical formula for the amplitude at $Q^2=0$ is obtained.
%; its $Q^2$ dependence at small but non-zero $Q^2$ is studied numerically.
We also estimate the shift of the muonic hydrogen energy levels, induced by $\sigma$ meson exchange. %such diagram;
For the $2S$ level the shift is approximately 30 times smaller than needed to resolve the proton radius puzzle,
but still exceeds the precision of the muonic hydrogen measurements.
\end{abstract}

\section{Introduction}

Two-photon exchange (TPE) in elastic lepton-proton scattering received much attention in the literature over last decade.
It was found that TPE could resolve the discrepancy between Rosenbluth and polarization transfer measurements of the proton form factors \cite{RP};
TPE effects are seen in low-$Q^2$ electron-proton scattering \cite{lowQ2} and in dedicated electron/positron scattering experiments \cite{VEPP}.
For a recent review see e.g. Ref.~\cite{TPErev}.

\begin{figure}
\centering
\includegraphics[width=0.2\textwidth]{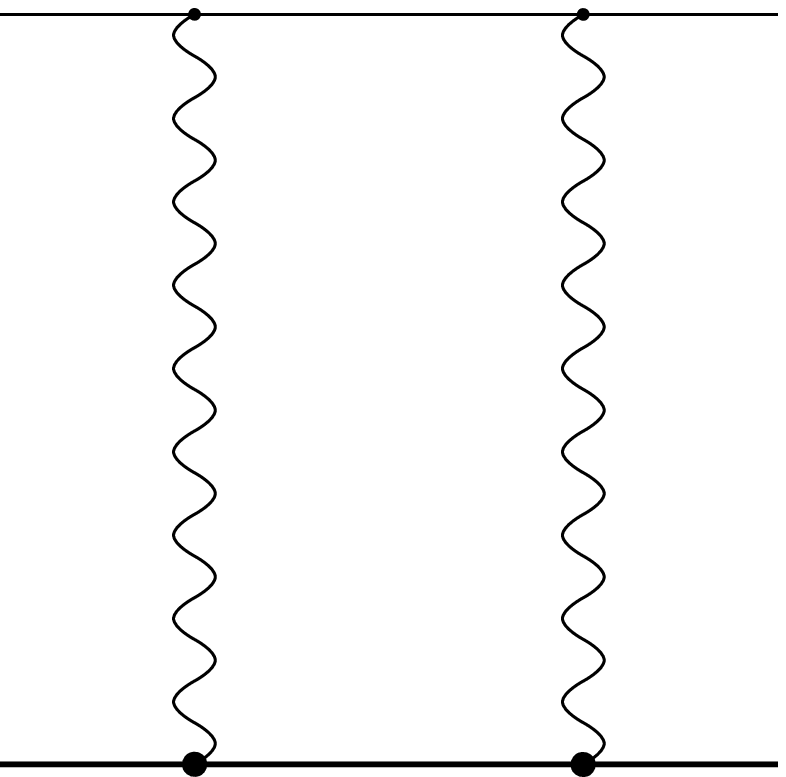}\hfil
\includegraphics[width=0.2\textwidth]{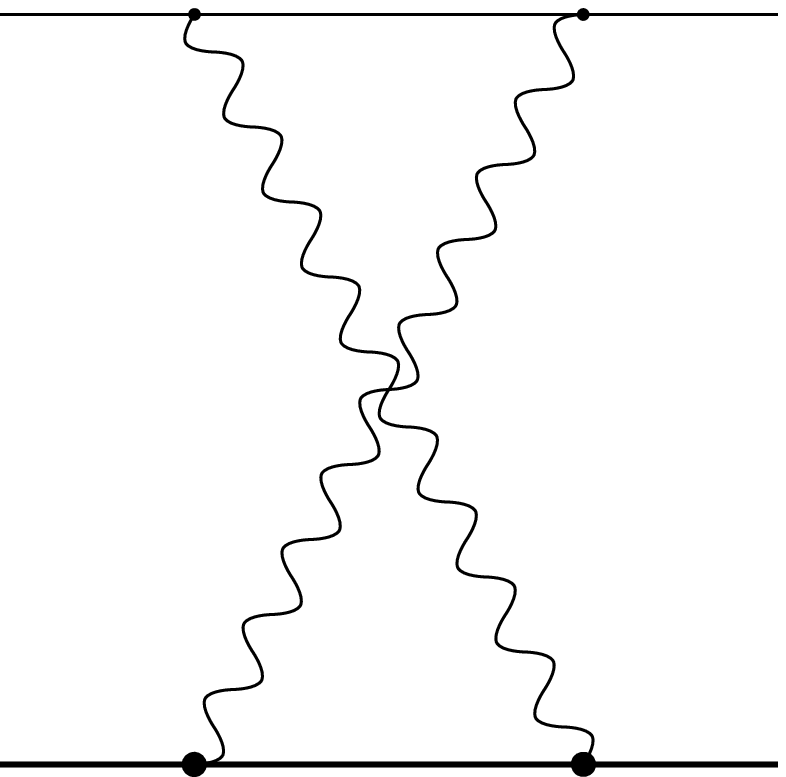}\hfil
\includegraphics[width=0.2\textwidth]{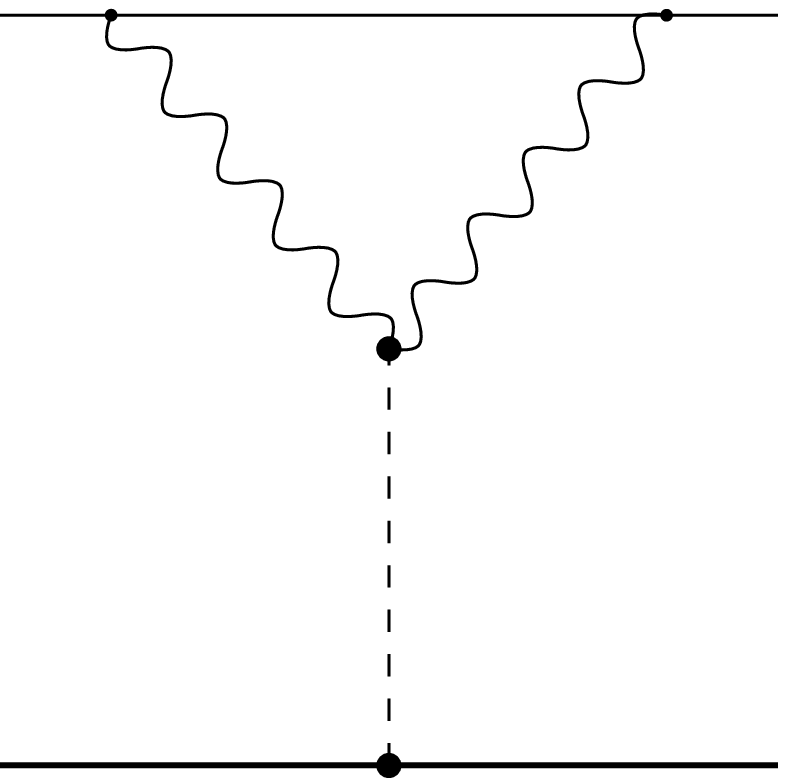}\\
(a)\hspace{0.275\textwidth}(b)\hspace{0.275\textwidth}(c)
\caption{Box, crossed-box and $\sigma$ meson exchange diagrams}\label{diagr}
\end{figure}

Usually, TPE is understood as a non-trivial second-order contribution to the scattering amplitude, with structure different from first Born approximation,
and is represented by box and crossed-box diagrams with various intermediate states, Fig.\ref{diagr}(a,b).
Another, completely  different, type of the two-photon contribution arises from the exchange of a single $C$-even meson,
which then decays into two photons, Fig.\ref{diagr}(c).
Such diagrams were considered in Ref.~\cite{Zhou} and, recently, in Ref.~\cite{1}.
It happens that the $\pi$ meson contribution is exactly zero for the unpolarized particles, and the authors of Ref.~\cite{1} conclude
that the primary candidate for the exchanged particle is $\sigma$ meson.
Such contribution is proportional to the lepton mass and is thus very different for electron and muon scattering;
it was shown in Ref.~\cite{1} that in the latter case it can reach 0.1\% in the kinematics of the MUSE experiment \cite{MUSE}.

Results presented in Ref.~\cite{1} suggest that the meson exchange contribution strongly grows at $Q^2\to 0$,
which may be interesting in connection with so-called "proton radius puzzle".
The values of the proton charge radius, obtained from the elastic electron-proton scattering
and from the measurements of the Lamb shift in muonic hydrogen, are in serious disagreement \cite{prp}.
Potentially, any effect, which makes a difference between muon and electron,
could be responsible for the discrepancy and should be checked; the meson exchange is one of them.

In the present paper, we consider $\sigma$ meson exchange in the electron-proton scattering,
study the behaviour of the corresponding amplitude at small $Q^2$ both analytically and numerically,
and estimate its effect on the (muonic) hydrogen energy levels and Lamb shift proton radius measurements.
In general, we use ideology similar to Ref.~\cite{1}, but try to obtain analytical results where possible.

\section{Analytical evaluation}

The amplitude corresponding to the diagram Fig.\ref{diagr}(c) is \cite{1}
\be \label{M}
 {\cal M} = \frac{g_{\sigma pp}}{q^2-m_\sigma^2} f_s(q^2) \, \bar u'u \, \bar U'U
\ee
where $m_\sigma$ is $\sigma$ meson mass, $u$ ($u'$) and $U$ ($U'$) are lepton and proton spinors,
$g_{\sigma pp}$ is $\sigma$-meson-proton interaction constant, and $f_s$ is a form factor,
arising from the upper (triangle) part of the diagram.

The $f_s$ form factor is the main unknown in the Eq.(\ref{M}), and we will calculate it below. It is determined by \cite{1}:
%  We start from the Eq.~(18) of Ref.~\cite{1}.
\be
  T = \bar u' u \; f_s(q^2) = i (4\pi\alpha)^2 \int \frac{d^4k''}{(2\pi)^4}
      \; \bar u' \frac{\gamma_\nu(\hat k'' + m)\gamma_\mu}{k''^2-m^2} u \;
      \frac{\Delta_{\mu\nu}}{q_1^2 q_2^2} 
\ee
where $\alpha$ is fine structure constant, $m$ is lepton mass, $\Delta_{\mu\nu}$ is $\sigma\gamma\gamma$ vertex function,
\be
  \Delta_{\mu\nu} = A (g_{\mu\nu} \; q_1 q_2 - q_{1\nu} q_{2\mu}) + B (q_1^2 \; q_{2\mu} - q_1 q_2 \; q_{1\mu}) (q_2^2 \; q_{1\nu} - q_1 q_2 \; q_{2\nu})
\ee
and $A$ and $B$ are two scalar form factors which depend on three variables: $q^2$, $q_1^2$ and $q_2^2$. The kinematics is:
\be
  q_1 = k''-k, \qquad q_2 = k'-k'', \qquad  q = k'-k = q_1+q_2
\ee
where $k$ ($k'$) are initial (final) lepton momenta, $q^2$ is momentum transfer squared and as usual $-q^2 = Q^2 > 0$.
After algebraic transformations with gamma matrices we have
\be
\begin{split}
  T = i (4\pi\alpha)^2 \int \frac{d^4k''}{(2\pi)^4} \; \frac{1}{k''^2-m^2} \frac{1}{q_1^2 q_2^2} \; \bar u' & \left\{  A [ (\hat k'' - m)(k''^2-m^2+q_1^2+q_2^2) + m(q^2-q_1^2-q_2^2) ] \right. \\
                      & \left. - B (\hat k'' - m) [(k''^2-m^2) \; qq_1 \; qq_2 + q^2 q_1^2 q_2^2 ] \right\} u
\end{split}
\ee
Then we use the fact that, for a symmetric function $f$,
\be
  \int \bar u'(\hat k'' - m)u \; f(q_1^2,q_2^2) d^4k'' = 
  \int \bar u' \hat q_1 u \; f(q_1^2,q_2^2) d^4k'' =
  \int \bar u' \frac {\hat q_1+ \hat q_2}{2} u \; f(q_1^2,q_2^2) d^4k'' \equiv 0
\ee
and
\be
  \int k''_\mu \frac{d^4k''}{k''^2-m^2} f(q_1^2,q_2^2)  =
  \int \left( \frac{k''k_+}{k_+^2} k_{+\mu} + q_\mu \cdot {\rm const}\right) \frac{d^4k''}{k''^2-m^2} f(q_1^2,q_2^2),
\ee
($k_+ = k+k'$ and the second term vanishes after contraction with $\bar u'$ and $u$), therefore
\be
  \int \bar u' \frac{\hat k'' - m}{k''^2-m^2} u \; f(q_1^2,q_2^2) d^4 k'' =
  \frac{m}{4m^2-q^2} \; \bar u'u \int \left( 2 + \frac{q^2-q_1^2-q_2^2}{k''^2-m^2} \right) d^4 k''
\ee

Using the above equations, we obtain
\be
  f_s = i (4\pi\alpha)^2 \frac{m}{4m^2-q^2} \int \frac{d^4k''}{(2\pi)^4}
        \left\{  A \left[\frac{2}{q_1^2} + \frac{2}{q_2^2} - \frac{4m^2+q_1^2+q_2^2-q^2}{q_1^2 q_2^2 (k''^2-m^2)}(q_1^2+q_2^2-q^2) \right]
                -B q^2 \left[ 2 - \frac{q_1^2+q_2^2-q^2}{k''^2-m^2} \right] \right\}
\ee
The form factor $B$ is not known experimentally, since it does not contribute to the cross-section when both photons are real.
Nevertheless, we see that in our equation it comes with a coefficient, proportional to $q^2$. Thus, at least at small $Q^2$ its contribution will be small.
This justifies using the approach similar to Ref.~\cite{1}: we will put $B=0$. The other form factor $A$ is modeled assuming vector meson dominance \cite{1}
\be
  A = \frac{g_{\gamma\gamma\sigma}}{\left(1-q_1^2/\la^2\right)\left(1-q_2^2/\la^2\right)}
\ee
where $g_{\gamma\gamma\sigma}$ is interaction constant,
which can be determined from the $\sigma\to\gamma\gamma$ partial width, and $\la$ is $\rho$ meson mass.
After this
\be
  f_s = - \frac{\alpha^2}{i\pi^2} \frac{m}{4m^2-q^2} \; g_{\gamma\gamma\sigma} \int d^4 k'' \frac{\la^4}{(\la^2-q_1^2)(\la^2-q_2^2)}
        \left[ \frac{2}{q_1^2} + \frac{2}{q_2^2} - \frac{4m^2+q_1^2+q_2^2-q^2}{q_1^2 q_2^2 (k''^2-m^2)}(q_1^2+q_2^2-q^2) \right]
\ee

With the help of trivial identity
\be
  \frac{\la^2}{\la^2-q^2} \; \frac{1}{q^2} = \frac{1}{\la^2-q^2} + \frac{1}{q^2}
\ee
we can express $f_s$ via integrals of the three following types
\be
  I_3(\mu, \la) = \frac{1}{i\pi^2} \int \frac{d^4 k''}{(q_1^2-\mu^2)(q_2^2-\la^2)(k''^2-m^2)},
\ee
\be
  I_2(\mu, \la) = \frac{1}{i\pi^2} \int \frac{d^4 k''}{(q_1^2-\mu^2)(q_2^2-\la^2)},
\ee
\be
  I_{2x}(\mu) = \frac{1}{i\pi^2} \int \frac{d^4 k''}{(q_1^2-\mu^2)(k''^2-m^2)}
\ee
and the result is
\be \label{f_s_final}
\begin{split}
  f_s = -\alpha^2 g_{\gamma\gamma\sigma} \frac{m}{4m^2-q^2}
              & \left\{ (q^2-2\la^2)(4m^2-q^2+2\la^2) \cdot I_3(\la, \la) \right. - \\
              & - (q^2-\la^2)(4m^2-q^2+\la^2) \cdot 2 I_3(0, \la) + \\
              & + q^2 (4m^2-q^2) \cdot I_3(0, 0) + \\
              & + 4 \la^2 \left[ I_2(\la, \la) - I_2(0, \la) \right] - \\
              & \left. - 2 \la^2 \left[ I_{2x}(\la) - I_{2x}(0) \right]
        \right\}
\end{split}
\ee
Some analytical expressions for the integrals $I_3$, $I_2$, $I_{2x}$ are given in Appendix \ref{app:integrals}.

\section{$Q \to 0$ limit}
The limiting case $Q \to 0$ (or more precisely, $Q \ll m, \la$) is of special interest because of its connection with the proton radius puzzle:
the typical lepton momenta in hydrogen-like atoms are $Q \sim \alpha m \ll m$.

Using equations from the Appendix \ref{app:integrals}, we obtain for $Q \to 0$:
\be \label{Q0start}
  I_3(\la,\la) = \frac{1}{2m^2} \left\{ \ln \frac{m^2}{\la^2}
     + \frac{1-\frac{2m^2}{\la^2}}{\sqrt{1-\frac{4m^2}{\la^2}}} \ln \frac{1+\sqrt{1-\frac{4m^2}{\la^2}}}{1-\sqrt{1-\frac{4m^2}{\la^2}}} \right\}
\ee
\be
  I_3(0,\la) = \frac{1}{2m^2} \left\{ \ln \frac{m^2}{\la^2}
     + \sqrt{1-\frac{4m^2}{\la^2}} \ln \frac{1+\sqrt{1-\frac{4m^2}{\la^2}}}{1-\sqrt{1-\frac{4m^2}{\la^2}}} \right\}
\ee
\be
  I_3(0,0) = -\frac{\pi^2}{2mQ}
\ee
%\be
%  I_2(\la,\la) - I_2(0,\la) = \frac{Q^2+\la^2}{Q^2} \ln \frac{Q^2+\la^2}{\la^2} + \frac{\sqrt{4\la^2+Q^2}}{Q} \ln \frac{\sqrt{4\la^2+Q^2}-Q}{\sqrt{4\la^2+Q^2}+Q}
%\ee
\be \label{Q0end}
  I_2(\la,\la) - I_2(0,\la) = -1
\ee

Inserting Eqs.(\ref{Q0start}-\ref{Q0end}) in Eq.~(\ref{f_s_final}) and keeping only leading terms in $Q$, we obtain
\be \label{fs0}
  f_s = -\frac{\alpha^2 g_{\gamma\gamma\sigma}}{2m^3} \left\{ \frac{\la^4 - 2m^2\la^2 + 4m^4}{\sqrt{1-\frac{4m^2}{\la^2}}} \ln \frac{1-\sqrt{1-\frac{4m^2}{\la^2}}}{1+\sqrt{1-\frac{4m^2}{\la^2}}}
    - \la^4 \ln \frac{m^2}{\la^2} - 2\la^2 m^2 \right\}
\ee
First, we see that contrary to visual impression from Figs.3-4 of Ref.~\cite{1}, $f_s$ is finite at $Q \to 0$.
Second, the term $2m^3$ in the denominator suggests the divergence at $m \to 0$. However, in fact there is no divergence:
an accurate calculation reveals that in this limit the previous equation turns to
\be
  f_s = - 3 m \alpha^2 g_{\gamma\gamma\sigma} \left( \frac{1}{2} + \ln \frac{m^2}{\la^2} \right)
\ee
which vanish at $m \to 0$.
Therefore, the effect will be much smaller for the electron than for the muon, as expected.

\section{Numerical evaluation}

\begin{figure}
\centering
\includegraphics[width=0.48\textwidth]{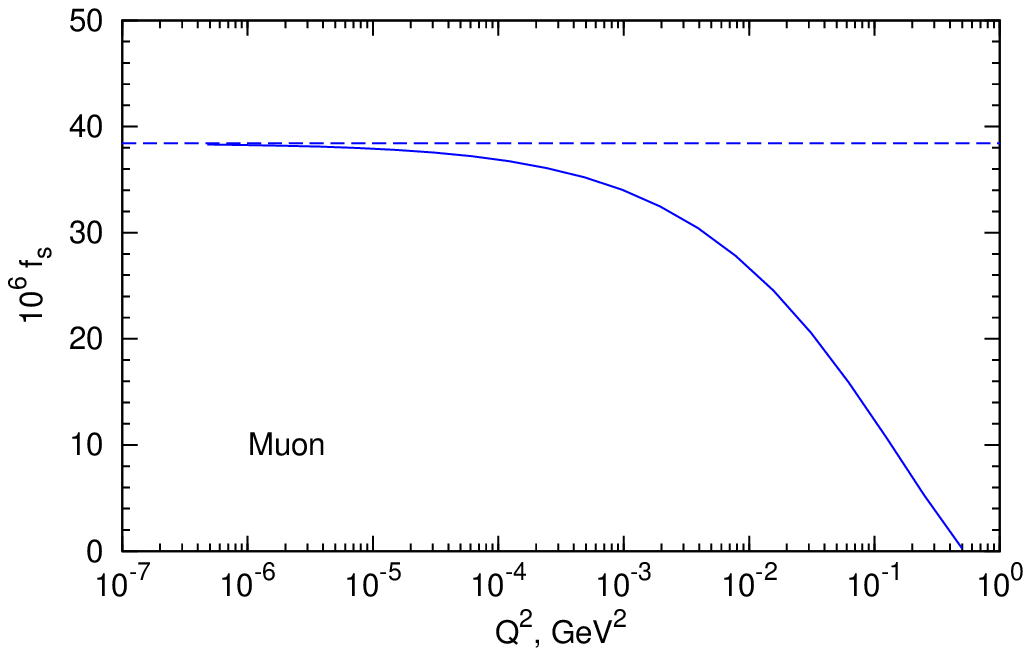}\hfil
\includegraphics[width=0.48\textwidth]{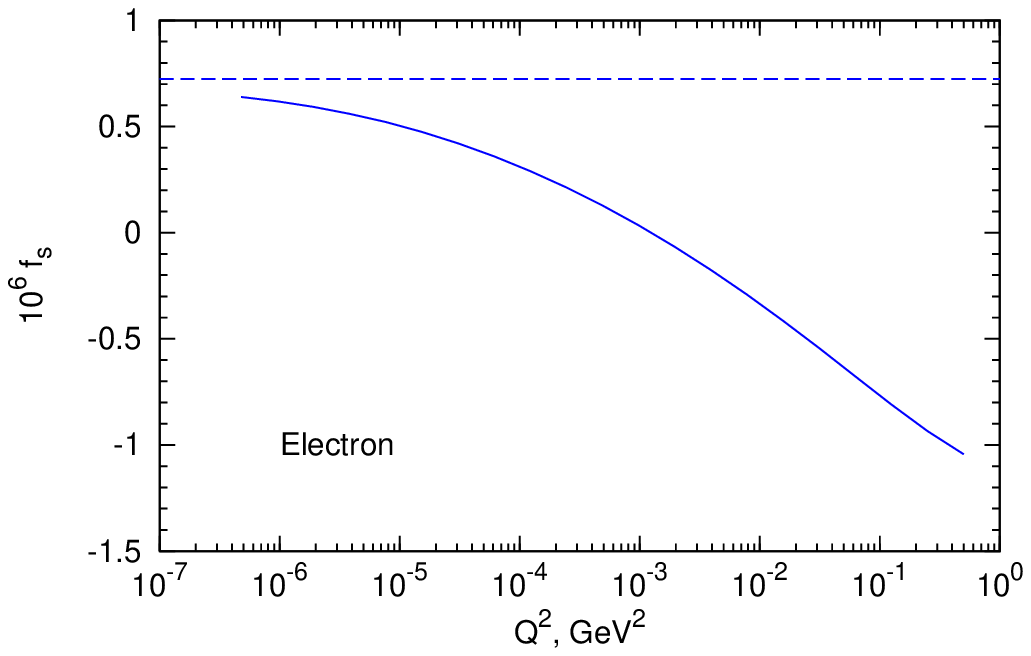}
\caption{$Q^2$ dependence of calculated $f_s$ form factor for muon and electron scattering.}\label{results}
\end{figure}

In our calculations we use the following values of the parameters: $g_{\sigma pp} = 5$, $m_{\sigma} = 0.5 \GeV$, $g_{\gamma\gamma\sigma} = 0.63 \GeV^{-1}$.
In Fig.\ref{results} we show $f_s(Q^2)$, calculated according to Eq.~(\ref{f_s_final}) for the electron and muon scattering.
Horizontal lines correspond to $Q^2 = 0$ limit according to Eq.~(\ref{fs0}):
\be
f_s(0) \approx 7.23\cdot 10^{-7} \text{ (electron), \qquad} f_s(0) \approx 3.84\cdot 10^{-5} \text{ (muon)}.
\ee
We also have calculated $f_s$ for the kinematical conditions considered in Ref.~\cite{1} and made sure that the results are consistent.

\begin{figure}
\centering
\includegraphics[width=0.48\textwidth]{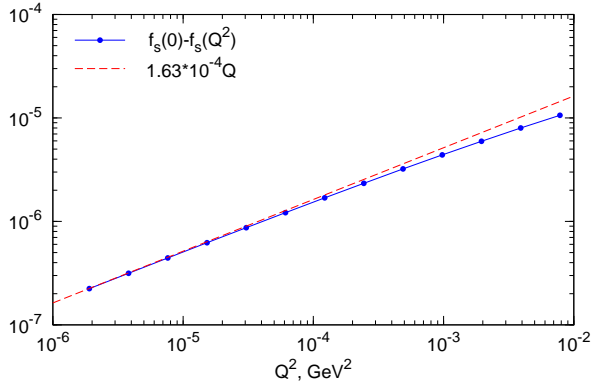}
\caption{$f_s(Q^2)$ for muon scattering and very small $Q^2$.}\label{precise}
\end{figure}

So, $f_s$ remains finite at $Q^2 \to 0$. How could be its behaviour at small $Q^2$ described more precisely?
We answer this question using simple numerical analysis.
In Fig.\ref{precise} we plot $f_s(Q^2) - f_s(0)$ for muon-proton scattering at small $Q^2$ in doubly logarithmic scale.
It can be seen from the plot that the points follow the straight line with the slope 1/2,
thus the dependence is $f_s(Q^2) = a + bQ$, where $a = f_s(0)$ as given by Eq.(\ref{fs0}) and $b \approx -1.63\cdot10^{-4} \GeV ^{-1/2}$.
The form factor is nonanalytical at $Q^2 = 0$, but this is not surprising:
it is known that the first singularity of a form factor corresponds to the lowest possible intermediate state mass; and for two photons this is zero.

\section{Implications for the proton radius puzzle}

Now we will calculate the shift of the energy levels of muonic hydrogen,
induced by the $\sigma$ meson exchange discussed in the previous section.

The scattering amplitude (\ref{M}) corresponds, in the nonrelativistic limit, to the additional potential energy
\be
 \delta U(r) = -\frac{1}{(2\pi)^3} \int g(Q^2) e^{i\vec Q \vec r} d\vec Q
\ee
where
\be \label{potential}
 g(Q^2) = - \frac{g_{\sigma pp}}{m_\sigma^2} f_s(Q^2)
\ee
The induced energy shift is (in the first-order perturbation theory)
\be \label{dEint}
 \Delta E = \int \delta U(r) |\psi(\vec r)|^2 d\vec r = - \frac{1}{2\pi^2} \int g(Q^2) \phi(r) \, Qr \sin Qr \, dr \, dQ
\ee
where $\phi(r)$ is radial density: $\phi(r) = \int |\psi(\vec r)|^2 d\Omega$, normalized as $\int\phi(r) r^2 dr = 1$.
Obviously the shift is non-zero only for spherically symmetric ($S$-wave) states.

Now let's calculate the energy shift in the case $g(Q^2) = a + bQ$. Since the shift is linear in $g$, we will have $\Delta E = \Delta E_a + \Delta E_b$,
where $\Delta E_a$ is proportional to $a$ and $\Delta E_b$ --- to $b$. Taking $g(Q^2) = a$ we easily obtain known result %\cite{dEcite}
\be
 \Delta E_a = -\frac{a}{4\pi} \phi(0)
\ee
For $g(Q^2) = bQ$ the integral (\ref{dEint}) diverges logarithmically at the upper limit in $Q$.
Since the dependence $g(Q^2) = a + bQ$ is actually an expansion at small $Q$ ($Q\ll m$),
it is natural to cut the integral at $Q = m$, still assuming that $m$ is large
compared to the characteristic scale of the wavefunction.
This way we can obtain (see Appendix \ref{app:shift})
\be \label{DeltaE_b}
 \Delta E_b = \frac{b}{\pi^2} \left\{ (\ln m\rho -1/2) \phi'(0)
          + \int_0^\infty \frac{dr}{r} \left[ \phi'(r) - \phi'(0) e^{-r/\rho} \right] \right\}
\ee
($\rho$ is arbitrary; in actual calculations it is convenient to put $\rho=$ Bohr radius of the corresponding orbit).
The totals for the $1S$ and $2S$ levels are
\bea
 \Delta E^{1S} = \Delta E_a^{1S} + \Delta E_b^{1S} &=& -\frac{(\alpha m)^3}{\pi} \left\{ a - \frac{8\alpha m}{\pi} \left( \ln 2\alpha + 1/2 \right) b \right\} \\
 \Delta E^{2S} = \Delta E_a^{2S} + \Delta E_b^{2S} &=& -\frac{(\alpha m)^3}{8\pi} \left\{ a - \frac{8\alpha m}{\pi} \left( \ln \alpha + 9/8 \right) b \right\}
\eea
Here $m$ should be reduced mass of the system, i.e. $m \approx 0.095 \GeV$ for the muonic hydrogen.
Using $a = -0.768\cdot 10^{-3} \GeV^{-2}$ and $b = 3.26\cdot 10^{-3} \GeV^{-3}$ (the values, determined in the previous section, should be multiplied by  $-g_{\sigma pp}/m_\sigma^2$ according to Eq.~(\ref{potential})), we obtain
\bea
 \Delta E_{1S} &=& 0.0814 - 0.0029 = 0.0785 \meV\\
 \Delta E_{2S} &=& 0.0102 - 0.0005 = 0.0097 \meV
\eea
(the first term is $\Delta E_a$, and the second is $\Delta E_b$).
Comparing this with the leading proton size contribution \cite{prp},
\be
 \Delta E_r = \frac{1}{6} \alpha r_E^2 \phi(0) \approx 5.20 \frac{r_E^2}{{\rm fm}^2} \meV
\ee
where $r_E$ is proton charge radius, we see that the $\sigma$ meson exchange leads to the shift of the measured radius
\be
  \Delta r_E = \frac{1}{2} r_E \frac{\Delta E^{2S}}{\Delta E_r} \approx 0.0012 {\rm\ fm}.
\ee
Taking into account values of the proton radius, obtained from the electron experiments: 0.8775(51) fm \cite{rE-electron}
and from the muon hydrogen measurements: 0.84087(39) fm \cite{rE-muon}, we see that $\Delta r_E$
is approximately 30 times smaller than the discrepancy between electron and proton results, but larger than the precision of the muonic measurements.

\section{Conclusions}

We studied two-photon contribution to the elastic lepton-proton scattering, associated with $\sigma$ meson exchange.
The effect is proportional to the lepton mass, and is therefore much more pronounced for the muon than for the electron.
Particular attention was paid to the low-$Q^2$ region.
We have shown that the corresponding amplitude grows sharply but remains finite at $Q^2\to 0$.
The analytical formulae for the amplitude at $Q^2= 0$ were obtained;
its $Q^2$ dependence at small but non-zero $Q^2$ was studied numerically and found to follow approximately $a+bQ$ law.

Then we estimate the shift of the (muonic) hydrogen energy levels, induced by $\sigma$ meson exchange; %such diagram;
in particular, we have found that for the $2S$ level the shift is $0.0097 \meV$.
In terms of the extracted proton radius the shift is about 0.0012 fm,
which is approximately 30 times smaller than needed to resolve the proton radius puzzle and goes in the wrong direction,
but still exceeds the precision of the muonic hydrogen measurements.
This calls for the careful examination of diagrams of this type,
which would allow to refine the value of the proton radius obtained from the Lamb shift measurements.

\section*{Acknowledgement}
The author is grateful to A.P.~Kobushkin for reading the manuscript and making useful comments.

\appendix
\section{Analytical formulae for the integrals}\label{app:integrals}

Using Feynman parameterization method, and after some easy transformations, the general integral $I_3$ can be represented as
\be
  I_3(\mu,\la) = \int_0^1 \frac{d\xi}{\sqrt{K(\xi)}} \ln \frac{\sqrt{K(\xi)} - [Q^2 + \la^2 - \mu^2 - Q^2\xi]}{\sqrt{K(\xi)} + [Q^2 + \la^2 - \mu^2 - Q^2\xi]} + (\la \leftrightarrow \mu)
\ee
where
\be
  K(\xi) = (\la^2 + \mu^2 + Q^2 - Q^2\xi)^2 - 4\la^2\mu^2 + 4m^2Q^2\xi^2.
\ee
From this representation, various limiting cases (such as $Q \to 0$, $m \to 0$) can be easily deduced.
A special case $I_3(0,0)$ can be written in a compact analytical form:
\be
  I_3 (0,0) = \frac{4}{Q\sqrt{4m^2+Q^2}}
  \left[ F\left( \frac{1-y}{1+y} \right) - F\left( \frac{y+1}{y-1} \right)\right],
\ee
where
\be
  y = \frac{2m+\sqrt{4m^2+Q^2}}{Q}, \qquad F(x) = \int_0^x \frac{\ln(1+z)}{z} dz
\ee

The integral $I_2$ is
\be
  I_2(\mu,\la) = R - \frac{Q^2+\la^2-\mu^2}{2Q^2} \ln \frac{\mu^2}{Q^2}
  + \frac{1}{2Q^2} \sqrt{K(0)} \ln\frac{\sqrt{K(0)} - [Q^2 + \la^2 - \mu^2]}{\sqrt{K(0)} + [Q^2 + \la^2 - \mu^2]} + (\la \leftrightarrow \mu)
\ee
where $R$ is infinite constant.

The integral $I_{2x}$ can be calculated
\be
  I_{2x}(\mu) = 2 \ln \frac{2R}{m} - \frac{4\ln\frac{\mu}{m}}{1+\sqrt{1-\frac{4m^2}{\mu^2}}} 
               +\frac{\mu^2}{m^2} \sqrt{1-\frac{4m^2}{\mu^2}} \ln \frac{1+\sqrt{1-\frac{4m^2}{\mu^2}}}{2} \qquad (\mu > 2m)
\ee
where again $R$ is infinite constant. At $m \to 0$ this expression diverges logarithmically.
If $\mu < 2m$, put $\sqrt{1-\frac{4m^2}{\mu^2}} \to i \sqrt{\frac{4m^2}{\mu^2}-1}$; at $\mu = 0$ only the first term survives.

\section{Derivation of Eq.(\ref{DeltaE_b})}\label{app:shift}

Starting from the Eq.(\ref{dEint}), we have
\be
  \Delta E_b = - \frac{b}{2\pi^2} \int_0^\infty \phi(r) \, dr \int_0^m Q^2 r \sin Qr \, dQ
\ee
%The inner integral is
%\be
%  \int_0^m Q^2 r \sin Qr \, dQ = -\frac{d}{dr} \left[ m\cos mr + \frac{2}{r} (\cos mr -1) \right]
%\ee
Doing simple integration over $Q$ and integrating over $r$ by parts, we obtain
\be
  \int_0^\infty \phi(r) \, dr \int_0^m Q^2 r \sin Qr \, dQ = \int_0^\infty \left[ m \sin mr - \frac{2}{r} (1-\cos mr) \right] \phi'(r) dr
\ee
Then
\be \label{i1}
  \left. \int m\sin mr \,\phi'(r) dr =  - \cos mr \,\phi'(r) \right|_0^\infty + \int \cos mr \,\phi''(r) dr = \phi'(0)
\ee
(the integral with cosine vanishes at $m\to \infty$) and, identically
\be \label{i2}
  \int \frac{1-\cos mr}{r} \,\phi'(r) dr =
        \int \frac{dr}{r} \left[ \phi'(r) - \phi'(0) e^{-r/\rho} \right] 
        - \int \frac{dr}{r} \cos mr  \left[ \phi'(r) - \phi'(0) e^{-r/\rho}\right ] 
        + \phi'(0) \int \frac{1-\cos mr}{r} e^{-r/\rho} dr
\ee
Since the function $\frac{\phi'(r) - \phi'(0) e^{-r/\rho}}{r}$ is regular at $r \to 0$,
the first integral in the r.h.s. is convergent, and the second one vanishes at $m\to\infty$.
The third integral can be calculated exactly:
\be
\int_0^\infty \frac{1-\cos mr}{r} e^{-r/\rho} dr = \frac{1}{2} \ln \left( 1 + m^2\rho^2 \right) \approx \ln m\rho
\ee
Adding all parts together, we obtain Eq.~(\ref{DeltaE_b}).

\end{document}